\documentclass[english,journal=jctcce,manuscript=review]{achemso}
\usepackage[T1]{fontenc}
\usepackage[latin9]{inputenc}
\usepackage{refstyle}
\usepackage{mathrsfs}
\usepackage{amsmath}

\makeatletter

\title{Improved Grid Optimization and Fitting in Least Squares Tensor Hypercontraction}
\author{Devin A. Matthews}
\affiliation{Southern Methodist University, Dallas, TX 75275, USA}
\email{damatthews@smu.edu}
\AtBeginDocument{\providecommand\figref[1]{\ref{fig:#1}}}
\RS@ifundefined{subsecref}
  {\newref{subsec}{name = \RSsectxt}}
  {}
\RS@ifundefined{thmref}
  {\def\RSthmtxt{theorem~}\newref{thm}{name = \RSthmtxt}}
  {}
\RS@ifundefined{lemref}
  {\def\RSlemtxt{lemma~}\newref{lem}{name = \RSlemtxt}}
  {}

\usepackage{pgfplots}
\pgfplotsset{compat=1.8}
\usetikzlibrary{patterns}
\usetikzlibrary{plotmarks}
\usepackage{pgfplotstable}
\usepgfplotslibrary{groupplots}
\usepackage{algorithmicx,algpseudocode}
\allowdisplaybreaks

\makeatother

\usepackage{babel}
\begin{document}
\begin{abstract}
A new method for generating fitting grids for least-squares tensor
hypercontraction (LS-THC) is presented. This method draws inspiration
from the related interpolative separable density fitting (ISDF) technique,
but uses only a pivoted Cholesky decomposition of the metric matrix,
$S$, already computed as a matter of course in LS-THC. The size and
quality of the resulting grid is controlled by a user-defined cutoff
parameter and the size of the starting grid. Additionally, the Cholesky-based
method provides an alternative and possible more numerically stable
method for performing the least-squares fit. The quality of the grids
produced is evaluated for LS-DF-THC-MP2 calculations on retinal and
benzene, the former with a large starting grid and small cc-pVDZ basis
set, and the latter with a wide range of grids and basis sets. The
error and grid size is found to be well-controlled by either the cutoff
parameter (with a large starting grid) or the starting grid size (with
a tight cutoff) and highly predictable. The Cholesky-based method
is also able to generate unique grids tailored to different charge
distributions, for example the $(ab|$, $(ai|$, and $(ij|$ distributions
that arise in the molecular orbital integrals. While only the $(ai|$
grid directly affects the MP2 energy, the relative sizes of the other
grids are examined.
\end{abstract}

\section{Introduction}

Tensor hypercontraction (THC)\cite{hohensteinTensorHypercontractionDensity2012}
is a promising tensor factorization technique that is, in theory,
applicable to any wavefunction-based electronic structure method.
The original conception of THC approximately factorized the two-electron
atomic orbital (AO) integrals into a product of five matrices,
\begin{equation}
(\mu\nu|\rho\sigma)\approx\sum_{PQ}X_{\mu}^{P}X_{\nu}^{P}V_{PQ}X_{\rho}^{Q}X_{\sigma}^{Q}\label{eq:thc}
\end{equation}
The structure of this approximation is similar in structure to a double
pseudo-spectral decomposition,\cite{friesnerSolutionSelfconsistentField1985,ringnaldaPseudospectralHartreeFock1990,martinezPseudospectralFullConfiguration1992a,martinezPseudospectralMoLler1994,martinezPseudospectralMultireferenceSingle1995}
although the method(s) for determining the values of the collocation
matrices $X$ and core matrix $V$ differ from those used in pseudo-spectral
theory as do their numerical behavior. Even more closely related is
the semi-numeric chain-of-spheres exchange (COSX) method of Neese
et al.\cite{neeseEfficientApproximateParallel2009} The THC factorization,
in combination with a Laplace quadrature of the orbital energy denominators
was used to derive reduced-scaling approaches to MP2,\cite{hohensteinTensorHypercontractionDensity2012,parrishTensorHypercontractionII2012}
MP3,\cite{hohensteinTensorHypercontractionDensity2012} CC2,\cite{hohensteinQuarticScalingSecondorder2013,hohensteinTensorHypercontractionEquationofMotion2013}
and the expensive particle-particle ladder term in CCSD.\cite{parrishCommunicationAccelerationCoupled2014}
Recent work on the application of this factorization scheme to the
coupled cluster doubles amplitudes, $\hat{T}_{2}$,\cite{hohensteinCommunicationTensorHypercontraction2012}
as well as the two-electron integrals has also lead to reduced scaling
CC approaches such as Tensor-Structured CC\cite{schutskiTensorstructuredCoupledCluster2017}
as well as THC versions of CASPT2,\cite{songReducedScalingCASPT22018}
ppRPA,\cite{shenviTensorHypercontractedPpRPA2014} and p2RDM methods.\cite{shenviTensorHypercontractedParametric2013}
Parrish et al. also showed that accuracy could be improved by directly
fitting the molecular orbital (MO) integrals $(pq|rs)$ rather than
transforming the factorization of the AO integrals after the fact.\cite{parrishTensorHypercontractionII2012,parrishCommunicationAccelerationCoupled2014}

In the least-squares variant of THC (LS-THC),\cite{parrishTensorHypercontractionII2012}
a global solution to (\ref{eq:thc}) via non-linear optimization is
abandoned in favor of an ansatz in which the matrix $X$ is fixed
by the choice of a grid $\{x_{P}\}_{P=1}^{n_{P}}$ and weights $\omega_{P}$
such that $X_{\mu}^{P}=\omega_{P}\phi_{\mu}(x_{P})$. However, this
approach is complicated by two factors: first, the choice of the molecular
grid is critical---it must be large enough to accurately represent
the electron-electron interaction, but not so large that the cost
of the calculation balloons or numerical issues are encountered.\cite{parrishDiscreteVariableRepresentation2013,kokkilaschumacherTensorHypercontractionSecondOrder2015}
Second, the LS-THC procedure requires the inversion of a ``metric
matrix'', $S$, which must be handled with care. This matrix is generally
rank-deficient (and hence singular), such that a pseudoinverse must
be constructed rather than a conventional inverse.

Lu and Ying worked around these problems, as well as the $\mathscr{O}(n^{4})$
or $\mathscr{O}(n^{5})$ cost of building the $E$ matrix, in the
context of periodic calculations.\cite{luCompressionElectronRepulsion2015}
They developed a method which they term the Interpolative Separable
Density Fitting (ISDF), where a randomized QR factorization of the
joint collocation matrix $Y_{\mu\nu}^{P}=X_{\mu}^{P}X_{\nu}^{P}$
is used to build a set of auxiliary functions $\tilde{\phi}_{P}$
that define the Coulomb kernel $V_{PQ}=(\tilde{\phi}_{P}|\tilde{\phi}_{Q})$
. This approach proved fruitful for reducing the cost of calculating
Hartree-Fock exchange\cite{luCompressionElectronRepulsion2015} and
RPA\cite{luCubicScalingAlgorithms2017} correlation energies, but
the application of ISDF to molecular systems is far less straightforward
due to the difficulty of computing the singular integrals required
for the Coulomb kernel.

In this Letter, we present a technique that bridges the ``classic''
LS-THC method with ideas from ISDF to arrive at a technique that is
capable of automatically determining optimized (pruned) grids specific
to a given $(pq|$ molecular orbital charge distribution. The quality
and size of the generated grids are investigated as a function of
a user-defined threshold $\epsilon$ and the parent grid size. This
technique is shown to produce high-quality grids with a much smaller
number of grid points than in the original grid, and to scale effectively
to basis sets as large as cc-pV6Z. The procedure for determining the
pruned grids also leads to an alternate method for performing the
least-squares fit which may be more numerically stable.

\section{Theory}

The LS-THC procedure is characterized by a closed-form solution to
the THC fit,\cite{parrishTensorHypercontractionII2012}
\begin{align*}
V & =S^{-1}ES^{-1}\\
S_{P^{\prime}Q^{\prime}} & =\sum_{pq}X_{p}^{P^{\prime}}X_{q}^{P^{\prime}}X_{p}^{Q^{\prime}}X_{q}^{Q^{\prime}}
\end{align*}
where $X$ is the (parent) grid collocation matrix, here in the MO
basis. The fitting matrix $E$ may be determined in a number of ways;
the density fitting approximation\cite{feyereisenUseApproximateIntegrals1993a,kendallImpactResolutionIdentity1997}
is used in this work.

In the ISDF approach,\cite{luCompressionElectronRepulsion2015} the
density is fit by an \emph{implicit} auxiliary basis $\tilde{\phi}_{P}(x_{P^{\prime}})$
defined only at a set of grid points. The auxiliary functions are
determined by a randomized sampled QR procedure,
\begin{align*}
M & =\mathcal{S}\mathcal{P}Y,\quad M\Pi=QR
\end{align*}
where the joint collocation matrix $Y$ in the starting grid is stored
as $Y_{pq,P}$. $\mathcal{P}$ is a permutation or mixing matrix (e.g.
FFT). $\mathcal{S}$ is a selection matrix which retains a random
set of $rN$ rows, where $r$ is an oversampling parameter and $N$
is the number of MOs. Finally, $M\Pi=QR$ computes the QR decomposition
with column pivoting. The auxiliary basis functions are then formed
by selecting $n_{P}\le n_{P^{\prime}}$ such that $|R_{n_{P}+1,n_{P}+1}|<\epsilon|R_{1,1}|\le|R_{n_{P},n_{P}}|$
and computing,
\[
\tilde{X}=R_{1:n_{P},1:n_{P}}^{-1}R_{1:n_{P},:}\Pi^{-1}
\]
where $\tilde{X}_{PP^{\prime}}=\tilde{\phi}_{P}(x_{P^{\prime}})$.
Together with explicit Fourier-space integration of $V_{PQ}=(\tilde{\phi}_{P}|\tilde{\phi}_{Q})$
via $\tilde{X}$ this gives a THC-like decomposition.

Rather than apply ISDF directly to molecules, it is instead interesting
to draw further parallels between ISDF and LS-THC by examining the
$M$ matrix. In fact, there is a close link between this quantity
and the metric matrix, $S$,
\begin{align*}
M^{T}M & =Y^{T}\mathcal{P}^{T}\mathcal{S}^{T}\mathcal{S}\mathcal{P}Y\\
 & \approx Y^{T}Y=S
\end{align*}
Additionally, the QR decomposition of $M$ may also be converted to
a decomposition of $S$,
\begin{align*}
S\approx M^{T}M & =\Pi R^{T}Q^{T}QR\Pi^{-1}\\
 & =\Pi R^{T}R\Pi^{-1}
\end{align*}
Thus, the $R$ factor can equivalently be obtained by a pivoted Cholesky
decomposition.\cite{harbrechtLowrankApproximationPivoted2012} As
in ISDF, only the leading portion of $R$ (using the same cutoff criterion
and threshold $\epsilon$) is numerically relevant to the computation.
The selected rows of $R$ also define the pruned grid $\{x_{P}\}$
from the original grid $\{x_{P^{\prime}}\}$ and the pruned collocation
matrix $X_{p}^{P}$ from $X_{p}^{P^{\prime}}$. Using different combinations
of $X_{a}^{P^{\prime}}$ and $X_{i}^{P^{\prime}}$ we can build three
distinct $S$ matrices corresponding to the $(ab|$, $(ai|$, and
$(ij|$ distributions. Each of these results in a unique pruned grid.

The Cholesky decomposition procedure allows for several critical optimizations.
Because the Cholesky factorization proceeds incrementally, we can
compute successive rows of $R$ until the diagonal falls below the
threshold and then stop early. In contrast, one must generally compute
all of the eigenvalues during pseudoinversion. Similarly, while both
algorithms scale as $\mathscr{O}(n^{3})$ the Cholesky factorization
has a much lower constant factor than eigendecomposition. However,
the main benefit of this approach compared to pseudoinversion is that
pruning the grid leads to a reduction in the cost of all following
computations (building the $E$ matrix, fitting, and the THC computation).
The Cholesky decomposition is a necessary factor for this optimization,
as the leading eigenvectors of $S$ may be arbitrary linear combinations
of grid points. These non-local functions are then no longer suitable
for defining an auxiliary basis since they destroy the property $(\mu\nu P)=(\mu P)(\nu P)=X_{\mu}^{P}X_{\nu}^{P}$.
In addition to decreasing computational cost, the use of the Cholesky
factorization may also improve numerical stability in the least squares
fitting solution. Instead of explicitly computing $S^{-1}=R^{-1}R^{-T}$
in the solution of $V=S^{-1}ES^{-1}$, we may instead solve the system
of equations $SVS=R^{T}RVR^{T}R=E$ using four triangular solves and
successive back-substitution. For ill-conditioned matrices, forward
solves are generally preferable to explicit inversion where possible.
Full pseudocode for the Cholesky-based LS-THC fitting procedure is
given in the Supporting Information.

\section{Results}

Since the Cholesky procedure can select an optimal (in some sense)
sub-grid from the parent grid, a natural question to ask is, ``Given
a large enough starting grid, how does the pruned grid size and accuracy
depend on the cutoff parameter $\epsilon$?'' In order to address
this question, we have performed a series of LS-THC-DF-MP2 calculations
on all-trans retinal using the cc-pVDZ basis set\cite{dunningGaussianBasisSets1989}
and corresponding cc-pVDZ-RI auxiliary basis set.\cite{weigendEfficientUseCorrelation2002}
A large parent grid with 49527 total points (1011 points/atom, $n_{P^{\prime}}/n_{DF}=31.6$
where $n_{DF}$ is the number of auxiliary functions) was used (see
SI for details). In fact, this grid has $n_{P^{\prime}}>n_{v}n_{o}$
which, in theory, is enough to exactly fit the $(ai|bj)$ integrals.
An experimental implementation of LS-THC-DF-MP2 in a development version
of the CFOUR program package\cite{stantonCFOURCoupledClusterTechniques}
was used, and the DF-MP2 results were calculated by reconstruction
of the $(ai|bj)$ integrals from the transformed DF integrals follow
by a conventional MP2 calculation.

\begin{figure}
\thispagestyle{empty}
\begin{tikzpicture}
\begin{axis}[
samples=100,
scale=0.9,
ymin=1e-9, ymax=1e-1,
xmin=1e-6, xmax=1e-1,
xmode=log, ymode=log,
x dir=reverse,
grid=major,grid style={dotted,line width=0.2pt,draw=gray},
mark size=1.5pt,
max space between ticks=20,
xlabel={$\epsilon$},
tick pos=left,
ylabel={$|\Delta E| \;/\; \text{E}_\text{h}$}
]
\addplot[draw=black,mark=*] coordinates {(1.000000e-02^0.5,2.486422e-01) (1.995262e-03^0.5,4.687727e-02) (3.981072e-04^0.5,3.528923e-03) (7.943282e-05^0.5,4.477776e-04) (1.584893e-05^0.5,7.695534e-05) (3.162278e-06^0.5,7.836826e-06) (6.309573e-07^0.5,9.862329e-07) (1.258925e-07^0.5,4.625380e-07) (2.511886e-08^0.5,2.914970e-07) (5.011872e-09^0.5,8.231746e-08) (1.000000e-09^0.5,5.569280e-09) (1.995262e-10^0.5,7.339110e-09) (3.981072e-11^0.5,1.995630e-09) (7.943282e-12^0.5,1.698500e-08) (1.584893e-12^0.5,6.692530e-07) (3.162278e-13^0.5,2.575634e-05) (6.309573e-14^0.5,3.608705e-04) (1.258925e-14^0.5,6.084366e-01) (2.511886e-15^0.5,5.354373e+01) (5.011872e-16^0.5,3.672902e+05) (1.000000e-16^0.5,2.710772e+09)};
\addplot[draw=black,dashed] coordinates {(1e-1,1e-1) (1e-6,1e-11)};
\end{axis}
\end{tikzpicture}

\caption{\label{fig:Accuracy}Accuracy of the LS-THC-DF-MP2 approximation compared
to canonical DF-MP2 for all-trans retinal as a function of the Cholesky
cutoff parameter $\epsilon$. The dashed line shows the linear relation
$\Delta E=10\,\text{E}_{\text{h}}\times\epsilon^{2}$.}
\end{figure}
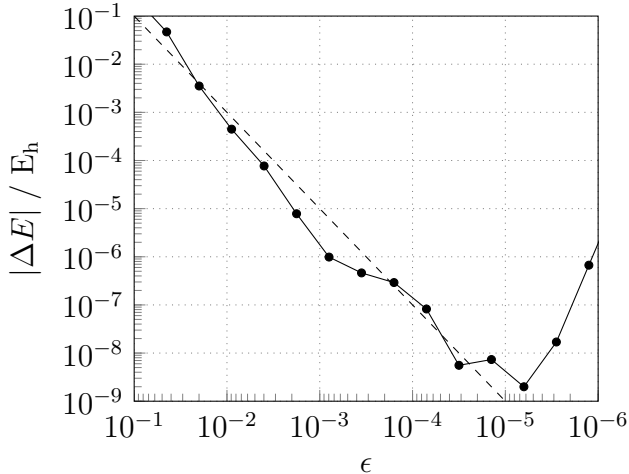

The error in the frozen-core LS-THC-DF-MP2 energy compared to canonical
DF-MP2\cite{kendallImpactResolutionIdentity1997} is illustrated in
\figref{Accuracy}. From these results we can see that the total error
decreases approximately linearly with the square of the cutoff $\epsilon$.
This is understandable since a cutoff of $\epsilon$ will disregard
a residual component of $S$ with diagonal elements at most $\epsilon^{2}\Vert S\Vert_{\max}$.
The magnitude of the residual diagonal elements is also closely related
to the magnitude of the eigenvalues of the rejected portion, such
that $\epsilon^{2}$ can be considered as roughly equivalent to the
cutoff used in the pseudo-inversion approach. Below $\epsilon\approx10^{-5}$,
the accuracy begins to degrade rapidly. This is likely due to numerical
stability issues encountered during the Cholesky decomposition or
the least-squares fitting. In some applications, the related $LDL^{T}$
decomposition can enhance numerical stability for positive semi-definite
problems, as it can capture the small negative diagonal elements that
spuriously arise due to round-off error. We intend to explore such
a decomposition in further work, although the achievable accuracy
seems to be entirely sufficient in this case.

The size of the pruned $(ab|$, $(ai|$, and $(ij|$ grids for different
choices of $\epsilon$ are highly linear w.r.t. $\log\epsilon$ below
$\epsilon=0.01$. The number of grid points per atom are closely fit
by $n_{P}^{(ab|}\approx-55\log\epsilon-21$, $n_{P}^{(ai|}\approx-37\log\epsilon-39$,
and $n_{P}^{(ij|}\approx-6\log\epsilon-6$, which gives $\sim$254,
146, and 24 points/atom at $\epsilon=10^{-5}$ respectively ($\sim8$,
4.5, and 0.75 $n_{DF}$). This $(ai|$ grid is of similar size to
the grids used in the original work on LS-THC which were hand-optimized
to reduce the error and required grid size. In this case, the tedious
hand optimization process is entirely replaced by an automated grid
optimization, with tunable error control.

The effect of the size of the starting grid on the pruned $(ai|$
grid and the relationship to the orbital basis set were investigated
using benzene, coupled with cc-pVXZ basis sets with $\text{X}=\text{D}$,
T, Q, 5, and 6 and their respective cc-pVXZ-RI auxiliary basis sets.
For each basis set, we performed calculations with 17 different parent
grids, ranging from relatively small (187 points/atom) to very large
(1648 points/atom); $\epsilon=10^{-5}$ was used throughout. For each
basis set, increasing the grid size eventually lead to a saturation
of the pruned grid, and a plateau in the error w.r.t. canonical DF-MP2.
While the largest grid is enough for an ``exact'' decomposition
even with cc-pV6Z, the saturated grids reliably prune $\sim60\%$
of the parent grid points, leading to between 115 and 615 points/atom.
The pruned grid size in units of $n_{DF}$ dropped from $\sim4.5$
at cc-pVTZ (cc-pVDZ is too close to the exact limit for a reasonable
comparison) down to $\sim3$ at cc-pV6Z. The saturated grids and pruning
fraction suggest reasonable sizes for the starting grid in the range
$7.5\le n_{P^{\prime}}/n_{DF}\le11$, although optional starting grid
optimization\cite{kokkilaschumacherTensorHypercontractionSecondOrder2015}
may also reduce starting grid size.

\section{Conclusions}

We have presented a modification of the LS-THC fitting procedure which
leads to automated pruning of the parent grid to grids specific to
the $(ab|$, $(ai|$, and $(ij|$ charge distributions. The extent
of pruning is controlled by the cutoff parameter $\epsilon$, the
size of the parent grid, and the orbital basis set employed. For a
large starting grid, the error is highly linear w.r.t $\epsilon$,
and the size of the pruned grids are also linear w.r.t. $\log\epsilon$.
As the size of the parent grid increases, the pruned grid quickly
reaches a saturation point, with final size (for the $(ai|$ distribution)
between $3n_{DF}$ and $4.5n_{DF}$. This technique shows promise
both as a method for in situ generation of optimal grids, and as a
tool for facilitating the definition of pre-generated grids for various
basis sets and levels of accuracy---in each case the ability to generate
distinct grids for different charge distributions is a novel characteristic.

\bibliography{paper}

\end{document}